\documentclass[10pt,prd,aps,a4paper,superscriptaddress,twocolumn,nofootinbib]{revtex4-2}

\usepackage{xspace} 
\usepackage{amsmath}
\usepackage{xcolor}
\usepackage{hyperref}
\usepackage{graphicx}
\usepackage{amssymb}
\usepackage{orcidlink}

\hypersetup{
    colorlinks=true,
    linkcolor=red,
    citecolor=blue,
    urlcolor=blue
}

\def\newacronym#1#2#3{\gdef#1{\gdef#1{#2\xspace}#3 (#2)\xspace}}

\newacronym{\bh}{BH}{black hole}
\newacronym{\gw}{GW}{gravitational wave}
\newacronym{\lvk}{LVK}{LIGO--Virgo--KAGRA}
\newacronym{\lisa}{LISA}{Interferometer Space Antenna}
\newacronym{\et}{ET}{Einstein Telescope}
\newacronym{\ce}{CE}{Cosmic Explorer}
\newacronym{\gr}{GR}{General Relativity}
\newacronym{\sm}{SM}{Standard Model of Particle Physics}
\newacronym{\dm}{DM}{dark matter}
\newacronym{\qnm}{QNM}{quasinormal mode}
\newacronym{\rw}{RW}{Regge-Wheeler}
\newacronym{\eos}{EOS}{equation of state}
\newacronym{\tov}{TOV}{Tolman-Oppenheimer-Volkoff}
\newacronym{\ec}{EC}{Einstein cluster}
\newacronym{\isco}{ISCO}{innermost stable circular orbit}
\newacronym{\emri}{EMRI}{extreme mass-ratio inspiral}
\newacronym{\dec}{DEC}{dominant energy condition}

\newcommand{\sapienza}{Dipartimento di Fisica, Sapienza Università 
	di Roma, Piazzale Aldo Moro 5, 00185, Roma, Italy}
\newcommand{\infn}{INFN, Sezione di Roma, Piazzale Aldo Moro 2, 00185, Roma, Italy}

\begin{document}

\title{Quasinormal modes and tidal responses of black holes\\ in generic anisotropic matter environments}

\author{Yu-Qian Zhao\orcidlink{0009-0000-8415-2367}}
\email[Yu-Qian Zhao: ]{yuqian.zhao@uniroma1.it}
\affiliation{\sapienza}
\affiliation{\infn}

\author{Paolo Pani\orcidlink{0000-0003-4443-1761}}
\email[Paolo Pani: ]{paolo.pani@uniroma1.it}
\affiliation{\sapienza}
\affiliation{\infn}

\begin{abstract}

We develop a perturbative framework for a black hole embedded in a generic, possibly anisotropic, matter environment under spherical symmetry.
Our approach extends previous analyses restricted to vanishing radial pressure or to perturbative matter configurations.
Within this framework, we derive an analytical generalization of the Einstein cluster that incorporates a polytropic radial pressure, and we investigate the properties of this solution.
We show that both the geodesic structure and the axial quasinormal-mode spectrum remain predominantly governed by an overall gravitational redshift effect, while the radial pressure systematically enhances the environmental corrections.
In contrast, the tidal Love numbers are substantially more sensitive, and can exhibit order-unity deviations, including vanishing and negative strictly static magnetic Love numbers for sufficiently large anisotropy.
We present the full linearized equations, which can be applied to various extensions, including ringdown analysis and extreme-mass-ratio inspirals.

\end{abstract}

\maketitle

\section{Introduction}

Since the first detection of the \gw event GW150914 in 2015~\cite{GW150914}, a global network of ground-based detectors, LIGO--Virgo--KAGRA~\cite{LIGO,VIRGO,KAGRA}, has established \gw astronomy as a new observational window on the Universe. 
The number of detected events has rapidly grown to $\mathcal{O}(10^2)$ over successive observing runs~\cite{GWTC1,GWTC2,GWTC3,GWTC4,GWTC5}, while future observatories such as LISA~\cite{LISA}, ET~\cite{ET:2025xjr}, and Cosmic Explorer~\cite{CE} are expected to significantly broaden the accessible frequency range and improve the precision of \gw measurements. 
As the sensitivity of current and next-generation detectors improves, it becomes increasingly important to assess how astrophysical environments surrounding compact objects may affect the generation and propagation of \gw{}s~\cite{Barausse:2014pra,Barausse:2014tra}.

In recent years, considerable progress has been made in developing relativistic and self-consistent descriptions of matter distributions surrounding \bh{}s within an effective-fluid framework, particularly in the context of the \ec~\cite{Cardoso:2021wlq,Cardoso:2022whc,Spieksma:2024voy,Chakraborty:2024gcr,Fernandes:2025osu}. Unlike traditional approaches based on Newtonian gravity or the slow-motion quadrupole approximation, this framework enables a fully relativistic treatment of environmental effects on \gw observables~\cite{Babak:2006uv,Destounis:2020kss,Destounis:2021rko,Zhao:2026yis}. It captures a broad class of environments, including \dm halos and axion clouds~\cite{Cardoso:2022whc}, and can be naturally extended to theories beyond GR~\cite{Fernandes:2025lon}.

The original \ec construction assumes vanishing radial pressure, an idealization that is mathematically convenient but physically restrictive. Even collisionless systems generically develop a small but finite radial pressure through velocity dispersion, which may leave observable imprints on \gw signals~\cite{binney2011galactic,Zhao:2026yis}. More generally, self-interactions, superfluid behavior, or large velocity dispersions can substantially enhance the effective radial pressure of the medium~\cite{Spergel:1999mh,Berezhiani:2025maf,Vogelsberger:2012ku}. Such effects may be particularly relevant in the strong-field regime, where \dm spike models predict significant density enhancements in the vicinity of \bh{}s~\cite{Gondolo:1999ef,Sadeghian:2013laa,Speeney:2022ryg,Mitra:2025tag,Zhao:2023tyo}. These considerations motivate extending the \ec paradigm to more general environments with nonvanishing radial pressure.

Recent work~\cite{Datta:2025ruh} initiated this program by developing an anisotropic perturbative formalism based on weak-environment expansions and multi-parameter perturbation theory. While such approaches successfully capture leading-order matter effects, they are intrinsically limited to small deviations from vacuum backgrounds~\cite{Dyson:2025dlj}. More recently, several studies have explored generalized perturbation schemes within the \ec framework~\cite{Chakraborty:2024gcr,DOnofrio:2026ulh,Rodriguez-Ruiz:2026xye}. However, these analyses retain the underlying \ec background and rely on simplified prescriptions for the fluid perturbations. A fully relativistic and self-consistent description of generic anisotropic environments with nonvanishing radial pressure is therefore still missing.

In this work, building on relativistic perturbation theory and the \ec formalism~\cite{chandrasekhar1985mathematical,Nagar:2005ea,Kojima:1992ie,Cardoso:2022whc,Chakraborty:2024gcr,DOnofrio:2026ulh}, we derive the complete set of perturbation equations describing the linear response of spherically symmetric \bh{}s embedded in generic astrophysical environments modeled as effective fluids. Our formulation consistently incorporates several limiting cases previously studied in the literature, including the \ec model, isotropic configurations, and weak-environment perturbative expansions, thereby providing a unified framework to investigate the combined effects of anisotropy and environmental strength.

Our primary goal is to elucidate the role of radial pressure and to explore its phenomenological consequences. To this end, we adopt a polytropic \eos for the radial pressure and derive a family of analytic background solutions by solving the Einstein equations in spherical symmetry, generalizing the original \ec construction~\cite{Cardoso:2021wlq}. We then investigate the corresponding energy conditions, which place theoretical constraints on the parameter space~\cite{Alho:2021sli,Datta:2023zmd}, and analyze the resulting geodesic structure, \qnm spectrum, and tidal response~\cite{Chakraborty:2026qru,Rodriguez:2026iot}. In particular, we study the properties of the light ring and the \isco, together with the associated \qnm{}s and tidal Love numbers, all of which play a central role in \gw generation and propagation~\cite{Ferrari:2020nzo,Muller:1999in,Nollert:1999ji,Berti:2009kk,Konoplya:2011qq,Berti:2025hly,Chakraborty:2026qru}.

We focus primarily on axial gravitational perturbations, which provide a particularly clean probe of the background geometry because they can be decoupled from matter fluctuations. Nevertheless, our formalism is fully general, and we also present the main equations governing the polar sector. We find that the dominant effect of the environment on axial perturbations is an overall gravitational redshift controlled by the environmental compactness, $\mathcal{C}$. This confirms the robustness of previous results obtained for environments without radial pressure~\cite{Cardoso:2021wlq,Cardoso:2022whc}, while revealing that radial pressure systematically amplifies the effect by replacing $\mathcal{C}$ with $(1+\mathcal{K})\mathcal{C}$, where $\mathcal{K}$ can be interpreted as the squared effective radial sound speed. Within the physically allowed parameter space, this enhancement can be of order unity relative to the \ec case, potentially increasing the observability of environmental signatures in \gw measurements and ringdown spectroscopy~\cite{Berti:2005ys,Spieksma:2024voy,Berti:2025hly}. More generally, our results demonstrate that radial pressure can have an impact comparable to, or even larger than, that of the matter distribution itself and should therefore be incorporated in realistic models of \bh environments.

Throughout this paper, we follow the notation of~\cite{Kojima:1992ie,Pani:2018inf,Ferrari:2020nzo,wald2010general} and adopt geometric units with $G=c=1$. Primes and overdots denote derivatives with respect to $r$ and $t$, respectively. Part of the calculations were performed using the \texttt{xAct} package~\cite{xAct}. Supplementary \texttt{Mathematica} notebooks containing definitions and intermediate results are publicly available at~\cite{MyGithub}.

\section{Framework}\label{sec:Framework}

We consider a static, spherically symmetric \bh immersed in a generic anisotropic environment. 
The background spacetime is described by the metric
\begin{equation}
ds^2=g_{\mu \nu}^{(0)}dx^\mu dx^\nu
=
-e^\nu dt^2+e^\lambda dr^2+r^2 d\Omega^2,
\end{equation}
where $d\Omega^2$ is the line element on the two-sphere, while $\nu(r)$ and $\lambda(r)$ are functions to be determined by the matter distribution surrounding the \bh.

The environment is modeled as an effective anisotropic fluid with stress-energy tensor~\cite{Raposo:2018rjn}
\begin{equation}
 T_{\mu \nu}^{\mathrm{env}(0)}
 =
 \rho u_\mu u_\nu
 +
 p_r k_\mu k_\nu
 +
 p_t \Pi_{\mu \nu},
\end{equation}
where $\rho$, $p_r$, and $p_t$ denote the effective density, radial pressure, and tangential pressure, respectively. 
For static configurations, these quantities depend only on the radial coordinate $r$. 
Here $u^\mu$ is the fluid four-velocity, while $k^\mu$ is a unit radial vector orthogonal to $u^\mu$, satisfying
$u_\mu u^\mu=-1$, $k_\mu k^\mu=1$, and $u_\mu k^\mu=0$. 
The tensor
\begin{equation}
\Pi_{\mu \nu}=g_{\mu \nu}+u_\mu u_\nu-k_\mu k_\nu
\end{equation}
projects onto the two-sphere orthogonal to both $u^\mu$ and $k^\mu$.

Including perturbations, the full metric can be decomposed as
\begin{equation}\label{eq:gmunuExpand}
g_{\mu \nu}
=
g_{\mu \nu}^{(0)}
+
g_{\mu \nu}^{\mathrm{axial}(1)}
+
g_{\mu \nu}^{\mathrm{polar}(1)},
\end{equation}
where the superscript $(1)$ denotes linear perturbations. 
As usual, perturbations split into axial (odd-parity) and polar (even-parity) sectors according to their behavior under parity transformations. 
The stress-energy tensor admits a similar expansion,
\begin{equation}\label{eq:TmunuExpand}
T_{\mu\nu}
=
T_{\mu \nu}^{\mathrm{env}(0)}
+
T_{\mu \nu}^{\mathrm{env}(1)},
\end{equation}
where $T_{\mu \nu}^{\mathrm{env}(1)}$ encodes the perturbation of the environment.

Following the standard \rw gauge decomposition~\cite{Regge:1957td,Zerilli:1970se,Nagar:2005ea}, the perturbations are expanded in scalar, vector, and tensor spherical harmonics. 
The metric perturbations are parametrized by the functions $(h_0^{\ell m}, h_1^{\ell m})$ in the axial sector and $(H_0^{\ell m}, H_1^{\ell m}, H_2^{\ell m}, K^{\ell m})$ in the polar sector. 
Matter perturbations are described by $(\delta \rho^{\ell m}, \delta p_r^{\ell m}, \delta p_t^{\ell m})$, together with the fluid variables $(R^{\ell m},V^{\ell m},U^{\ell m},Z^{\ell m},X^{\ell m})$. 
The explicit definitions are summarized in Appendix~\ref{app:Perturbation}.

The dynamics of the system follows from the Einstein equations and the covariant conservation of the stress-energy tensor,
\begin{align}\label{eq:Einstein}
\mathcal{E}_{\mu\nu}
=
G_{\mu\nu}
-
8\pi T_{\mu\nu}
&=
0,\\
\label{eq:DTmunu}
\nabla_\mu T^{\mu\nu}&=0,
\end{align}
where $G_{\mu\nu}$ is the Einstein tensor associated with the full metric $g_{\mu\nu}$. 
Substituting Eqs.~\eqref{eq:gmunuExpand} and \eqref{eq:TmunuExpand} yields the equations governing both the background and the perturbations.

Introducing the mass function $M(r)$ through
\begin{equation}
e^{-\lambda}=1-\frac{2M(r)}{r},
\end{equation}
the background equations reduce to the anisotropic \tov system~\cite{Doneva:2012rd},
\begin{align}\label{eq:tov1}
M^{\prime}&=4\pi r^2\rho,\\
\label{eq:tov2}
\nu^{\prime}
&=
\frac{2\left(M+4\pi r^3 p_r\right)}
{r\left(r-2M\right)},
\\
\label{eq:tov3}
p_r^{\prime}
&=
-\frac{\left(M+4\pi r^3 p_r\right)\left(\rho+p_r\right)}
{r\left(r-2M\right)}
-\frac{2\sigma}{r},
\end{align}
where the anisotropy parameter is defined as
\begin{equation}
\sigma(r)=p_r(r)-p_t(r).
\end{equation}
The system therefore consists of three equations for the five unknown functions $(M,\nu,\rho,p_r,p_t)$. 
Two additional relations must be specified in order to close the system, depending on the environmental model under consideration. 
These conditions will be introduced in Sec.~\ref{sec:Background}.

As discussed above, perturbations split into axial and polar sectors. 
The axial sector decouples from matter perturbations and can be reduced to a Schrödinger-like master equation,
\begin{equation}\label{eq:AxialTime}
\left(
\frac{\partial^2}{\partial r_*^2}
-
\frac{\partial^2}{\partial t^2}
-
V(r)
\right)\psi=0,
\end{equation}
where the tortoise coordinate is defined by
\begin{equation}
\frac{dr_*}{dr}=e^{(\lambda-\nu)/2}.
\end{equation}
The effective potential $V(r)$ depends on the fluid's properties. Indeed,
as discussed in~\cite{Pani:2018inf,Chakraborty:2024gcr,DOnofrio:2026ulh}, different choices for the fluid variables lead to distinct perturbation schemes. 
In this work, we focus on two physically motivated cases, namely the \textit{strictly static} and \textit{irrotational} configurations, which satisfy
\begin{equation}
U^{\ell m}=X^{\ell m}=0,
\end{equation}
and
\begin{equation}
U^{\ell m}=-(\rho+p_t)h_0^{\ell m},
\qquad
X^{\ell m}=-h_1^{\ell m},
\end{equation}
respectively. The latter choice corresponds to a fluid with zero vorticity~\cite{Landry:2015cva,Pani:2018inf}.

For these two cases, we obtain
\begin{equation}
V^{\mathrm{sta}}
=
e^\nu
\left[
\frac{\ell(\ell+1)}{r^2}
-
\frac{6M}{r^3}
+
4\pi(\rho-p_r-4\sigma)
\right],
\end{equation}
\begin{equation}
V^{\mathrm{irr}}
=
e^\nu
\left[
\frac{\ell(\ell+1)}{r^2}
-
\frac{6M}{r^3}
+
4\pi(\rho-p_r)
\right].
\end{equation}

Further details of the derivation are provided in Appendix~\ref{app:Perturbation}. 
Moreover, as shown in Appendix~\ref{app:Limits}, our formalism consistently reproduces the vacuum, isotropic, \ec, and weak-environment limits under the appropriate assumptions.

Polar perturbations are instead coupled to matter fluctuations and are therefore considerably more involved~\cite{Cardoso:2022whc}. 
For completeness, we summarize the corresponding equations in Appendix~\ref{app:Perturbation} and in the accompanying \texttt{Mathematica} notebook~\cite{MyGithub}. 
In the present work, however, we restrict our analysis to the axial sector.

\section{Black holes embedded in matter halos}\label{sec:Background}

Having established the general framework in Sec.~\ref{sec:Framework}, we now focus on the case of \bh{}s embedded in \dm halos, which provide a well-motivated and physically relevant realization of the generic environments discussed above.

Inspired by Refs.~\cite{Cardoso:2021wlq,Zhao:2023itk}, we adopt the same mass profile and model the radial pressure with a polytropic \eos,
\begin{align}
M(r)&=M_{\mathrm{BH}}+\frac{M_{\mathrm{DM}} r^2}{\left(a_0+r\right)^2}\left(1-\frac{2 M_{\mathrm{BH}}}{r}\right)^2,\\
    p_r&=\mathcal{K} \rho^\gamma,
\end{align}
where $M_{\mathrm{BH}}$ is the \bh{} mass, $M_{\mathrm{DM}}$ characterizes the total mass of the halo, and $a_0$ is its typical length scale.

Substituting these expressions into Eqs.~\eqref{eq:tov1} and \eqref{eq:tov3}, we obtain the density profile and tangential pressure,
\begin{equation}
    \rho=\frac{M_{\mathrm{DM}}(a_0+2 M_{\mathrm{BH}}) (r-2 M_{\mathrm{BH}})}{2 \pi  r^2 (a_0+r)^3},
\end{equation}
\begin{equation}\label{eq:pt}
p_t=p_r+\frac{r p_r^{\prime}}{2}+\frac{\left(M+ 4 \pi  r^3 p_r\right)\left(p_r+\rho\right)}{2\left(r-2 M\right)}.
\end{equation}

The metric function $g_{tt}=-e^{\nu(r)}$ can be obtained by integrating Eq.~\eqref{eq:tov2}. 
Analytical solutions can be found for several values of $\gamma$, in particular for $\gamma=1,2,3$. They take the form
\begin{equation}
    e^\nu=\left(1-\frac{2M_{\mathrm{BH}}}{r}\right)e^\Upsilon\zeta^\mathcal{K},
\end{equation}
\begin{equation}
   \Upsilon= \sqrt{\frac{M_{\mathrm{DM}}}{\xi}}\left[2  \arctan \frac{r+a_0-M_{\mathrm{DM}}}{\sqrt{M_{\mathrm{DM}} \xi}}-\pi \right],
\end{equation}
\begin{equation}
    \xi=2 a_0-M_{\mathrm{DM}}+4 M_{\mathrm{BH}},
\end{equation}
For $\gamma=1$, one finds
\begin{equation}
    \zeta=e^\Upsilon\frac{\left(r+a_0\right)^2}{(r+a_0)^2-2 M_{\mathrm{DM}}(r-2 M_{\mathrm{BH}})},
\end{equation}
These solutions naturally reduce to the \ec case~\cite{Cardoso:2021wlq} when $\mathcal{K}=0$. 
In the following, we focus on the case $\gamma=1$, which allows for analytical control and facilitates a direct comparison with previous results. 
For completeness, the analytical solutions for $\gamma=2$ and
$\gamma=3$ are given in~\cite{MyGithub}.
Since the mass function $M$ and the corresponding density profile $\rho$ are the same as in Ref.~\cite{Cardoso:2021wlq}, and the additional factor $\zeta^{\mathcal{K}}$ is positive and finite for $\xi>0$, the causal structure of the solution is not affected by the radial pressure.
For $M_{\mathrm{DM}}<2(a_0+2M_{\mathrm{BH}})$, i.e., $\xi>0$, the spacetime is asymptotically flat with ADM mass $M_{\mathrm{BH}}+M_{\mathrm{DM}}$, and has a single regular horizon at $r=2M_{\mathrm{BH}}$ and a curvature singularity at $r=0$, for any $\mathcal{K}$.
For $M_{\mathrm{DM}}\geq2(a_0+2M_{\mathrm{BH}})$, i.e., $\xi\leq0$, additional curvature singularities appear outside the horizon, which is far from the astrophysical one $M_{\mathrm{DM}}\ll a_0$ and out of our consideration.

\begin{figure}  \includegraphics[width=0.48\textwidth]{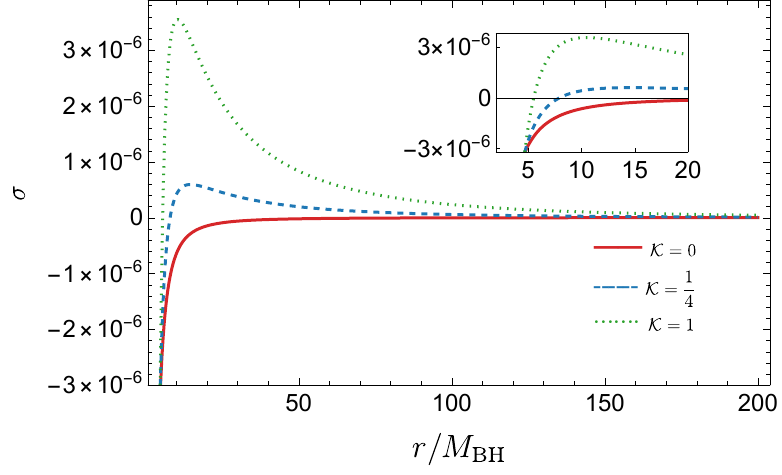}
    \caption{Anisotropy parameter $\sigma=p_r-p_t$ for different values of $\mathcal{K}$.
    The cases $\mathcal{K}=0$ and $\sigma=0$ correspond to the \ec and locally isotropic limits, respectively, while positive values correspond to radial-pressure-dominated anisotropic configurations.
    For illustration, we choose $M_{\mathrm{DM}}=10M_{\mathrm{BH}}$ and $a_0=100M_{\mathrm{BH}}$.
    }
    \label{fig:sigma}
\end{figure}

For the solutions considered here ($\gamma=1$), $\mathcal{K}$ can be directly interpreted as the square of the sound speed along the radial direction,
\begin{equation}
     c_{s_r}^2=\frac{\partial p_r}{\partial\rho}=\mathcal{K}.
\end{equation}
At the same time, $\mathcal{K}$ also controls the anisotropic structure through $\sigma=p_r-p_t$, as illustrated in Fig.~\ref{fig:sigma}.
In particular, the transition from $\sigma<0$ to $\sigma>0$ marks the regime where the radial pressure becomes dominant over the tangential component.
The limits $\mathcal{K}=0$ and $\sigma=0$ correspond to the \ec and locally isotropic cases, respectively.
In the former \ec case with $\mathcal{K}=0$, the matter consists of collisionless particles on circular orbits without any interaction~\cite{Cardoso:2021wlq}.
There is no radial motion, therefore, neither radial pressure nor radial propagation of sound waves.

The strong, weak, null, and dominant energy conditions read (see, e.g.,~\cite{Alho:2021sli})
\begin{equation}
\begin{aligned}
& \mathrm{SEC}: \rho + p_{ r} + 2 p_{t} \geq 0, \quad \rho + p_{ r} \geq 0, \quad \rho + p_{t} \geq 0, \\
& \mathrm{WEC}: \rho \geq 0, \quad \rho + p_{ r} \geq 0, \quad \rho + p_{t} \geq 0, \\
& \mathrm{NEC}: \rho + p_{ r} \geq 0, \quad \rho + p_{t} \geq 0, \\
& \mathrm{DEC}: \rho \geq |p_{r}|, \quad \rho \geq |p_{t}|.
\end{aligned}
\end{equation}
The effective density $\rho$, radial pressure $p_r$, and tangential pressure $p_t$ are found to satisfy the first three conditions everywhere. However,
the \dec can be violated in the vicinity of the \bh{} depending on the choice of parameters~\cite{Datta:2023zmd}. 

These violations are confined to the innermost region surrounding the \bh. This behavior is consistent with expectations from \dm spike models, in which the matter density is strongly depleted within a characteristic radius of $r\sim \mathcal{O}(4M_{\mathrm{BH}})$, or near the \isco, due to various dynamical processes~\cite{Gondolo:1999ef,Sadeghian:2013laa,Speeney:2022ryg,Mitra:2025tag,Zhao:2023tyo}. As a result, the region where the \dec is violated naturally coincides with the domain in which the effective-fluid description is expected to cease to be valid. From this perspective, such violations need not signal a fundamental inconsistency of the model, but rather reflect the limitations of the effective description in a regime where the underlying matter distribution is no longer accurately captured by a fluid approximation, as discussed in~\cite{Cardoso:2021wlq}.

From an observational standpoint, the \dec{} need only hold within the region relevant to the phenomenon being probed. In particular, the \qnm{} spectrum is mainly determined by the geometry in the vicinity of the light ring, while \emri observations are sensitive to a broader region extending further beyond the \isco.

\begin{figure}
    \centering
\includegraphics[width=0.48\textwidth]{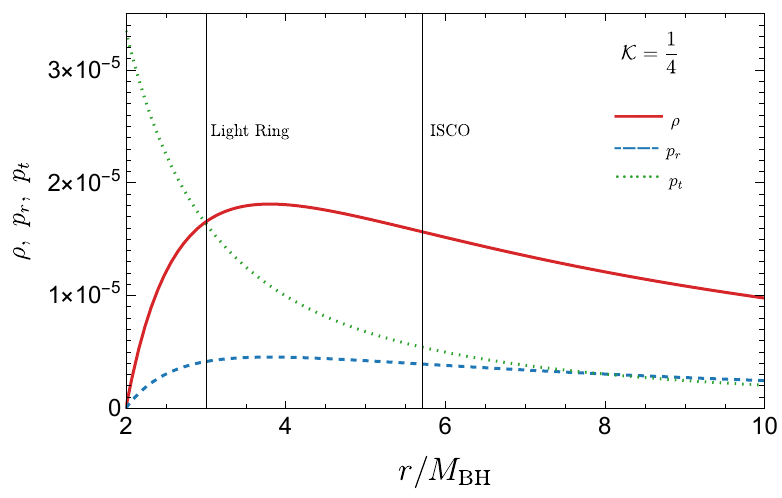}
\includegraphics[width=0.48\textwidth]{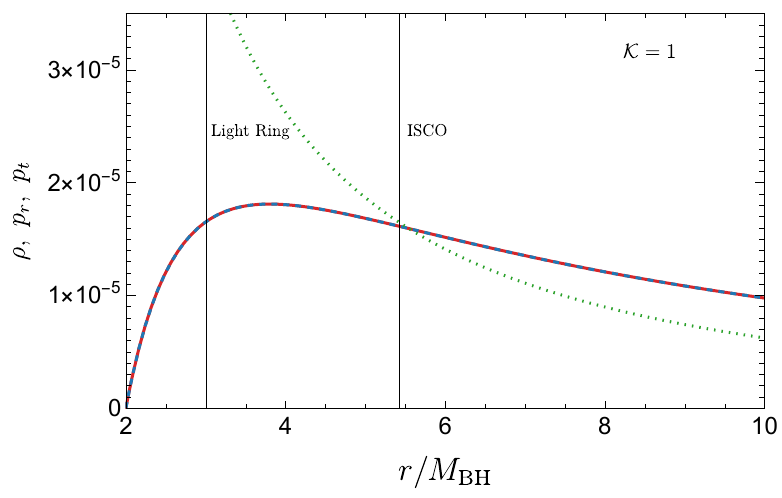}
   \caption{
Profiles of the effective density $\rho$, radial pressure $p_r$, and tangential pressure $p_t$, together with the corresponding energy conditions, for $\mathcal{K}=1/4$ and $\mathcal{K}=1$. The \dec{} is violated only in the innermost region, whose location depends only weakly on $M_{\mathrm{DM}}$ and $a_0$ in the realistic regime $\mathcal{C}=M_{\mathrm{DM}}/a_0\ll1$. The remaining parameters are the same as in Fig.~\ref{fig:sigma}.
}
    \label{fig:energy}
\end{figure}

Using Eq.~\eqref{eq:pt}, the onset of these violations can be determined analytically and is found to depend only weakly on $M_{\mathrm{DM}}$ and $a_0$ for realistic astrophysical configurations ($a_0 \gg M_{\mathrm{DM}}$). As illustrated in Fig.~\ref{fig:energy}, requiring that the \dec{} be satisfied up to the light-ring radius imposes the constraint
\begin{equation}\label{eq:K14}
\mathcal{K} \lesssim \frac{1}{4},
\end{equation}
whereas requiring it only up to the \isco{} yields
\begin{equation}\label{eq:SoundSpeed}
   \mathcal{K}\leq1,
\end{equation}
which is equivalent to the causality bound on the radial sound speed.

\section{Signatures of the model}

In this section, we investigate the observable consequences of our model. In particular, we analyze how radial pressure affects the geodesic structure, the \qnm spectrum, and the tidal Love numbers~\cite{Chakraborty:2026qru,Rodriguez:2026iot}, thereby probing the environment through complementary GW observables.

\subsection{Geodesic structure: light ring and ISCO}\label{sec:Geodesic}

The geodesic structure provides a direct probe of the strong-field region surrounding the \bh and plays a central role in a variety of \gw observables. In particular, unstable null geodesics determine the location of the light ring and are closely connected to the properties of the \bh shadow and to the eikonal limit of \qnm{}s~\cite{Cardoso:2008bp}. Timelike geodesics are equally important: the \isco marks the onset of orbital instability for massive particles and is therefore relevant for accretion processes, \emri systems, and for the dynamics of matter surrounding the \bh~\cite{Mitra:2025tag,Speeney:2022ryg}.

For realistic astrophysical environments, the compactness of the matter distribution is expected to satisfy $\mathcal{C}=M_{\mathrm{DM}}/a_0\lesssim10^{-4}$. Expanding the relevant quantities up to $\mathcal{O}(\mathcal{C}^2)$, the light-ring radius of the solution reads
\begin{equation}
r_{\mathrm{LR}}=3M_{\mathrm{BH}}\left[1+\left(1+2\mathcal{K}\right)\frac{M_{\mathrm{BH}}}{M_{\mathrm{DM}}}\mathcal{C}^2\right],
\end{equation}
while the corresponding angular frequency and Lyapunov exponent are~\cite{Cardoso:2008bp}
\begin{equation}\begin{aligned}
   & M_{\mathrm{BH}}\Omega_{\mathrm{LR}}=\frac{1}{3\sqrt{3}}\left\{1-\left(1+\mathcal{K}\right)\mathcal{C}\right.\\
    &\left.+\frac{1}{6}\left[\left(1+4 \mathcal{K}+3 \mathcal{K}^2\right) +6 (3+4\mathcal{K}) \frac{M_{\mathrm{BH}}}{M_{\mathrm{DM}}}\right] \mathcal{C}^2\right\},
\end{aligned}\end{equation}
\begin{equation}
    M_{\mathrm{BH}}\lambda=
    \frac{1}{3\sqrt{3}}\left[1-\left(1+\mathcal{K}\right)\mathcal{C}
    +\frac{1}{6}\left(1+4\mathcal{K}+3\mathcal{K}^2\right)\mathcal{C}^2\right].
\end{equation}
The critical impact parameter, which governs the capture of high-frequency photons and \gw{}s, is
\begin{equation}\begin{aligned}
   & b_{\mathrm{crit}}=3\sqrt{3}M_{\mathrm{BH}}\left\{1+\left(1+\mathcal{K}\right)\mathcal{C}\right.\\
    &\left.+\frac{1}{6}\left[\left(5+8 \mathcal{K}+3 \mathcal{K}^2\right) -6 (3+4\mathcal{K}) \frac{M_{\mathrm{BH}}}{M_{\mathrm{DM}}}\right] \mathcal{C}^2\right\}.
\end{aligned}\end{equation}

These results reveal a clear hierarchy. The leading correction, proportional to $\mathcal{C}$, is entirely due to the global gravitational redshift produced by the environment. Genuine modifications of the null-geodesic structure arise only at order $\mathcal{C}^2$, through changes in the effective potential governing photon trajectories. Radial pressure enhances both effects relative to the \ec case~\cite{Cardoso:2021wlq}.

As a consequence, the correction to the \bh shadow is suppressed by $\mathcal{O}(\mathcal{C}^2)\lesssim10^{-8}$ for realistic astrophysical environments. Environmental effects on current shadow observations are therefore expected to be negligible~\cite{EventHorizonTelescope:2019dse,Abuter:2018uum}.

The behavior of timelike geodesics follows a similar pattern. The \isco radius is
\begin{equation}
    r_{\mathrm{ISCO}}
    =6M_{\mathrm{BH}}
    \left[1-32\left(1+2\mathcal{K}\right)
    \frac{M_{\mathrm{BH}}}{M_{\mathrm{DM}}}\mathcal{C}^2\right],
\end{equation}
with no ${\cal O}(\mathcal{C})$ correction for any $\mathcal{K}$. The corresponding orbital frequency is
\begin{equation}\begin{aligned}
   & M_{\mathrm{BH}}\Omega_{\mathrm{ISCO}}=\frac{1}{6\sqrt{6}}\left\{1-\left(1+\mathcal{K}\right)\mathcal{C}\right.\\
    &\left.+\frac{1}{6}\left[\left(1+4 \mathcal{K}+3 \mathcal{K}^2\right) +12 (33+65\mathcal{K}) \frac{M_{\mathrm{BH}}}{M_{\mathrm{DM}}}\right]\mathcal{C}^2\right\}.
\end{aligned}\end{equation}

Both null and timelike geodesics therefore lead to the same conclusion: the dominant imprint of the anisotropic environment is a global ($\mathcal{K}$-dependent) redshift of characteristic frequencies, whereas the underlying orbital structure is only weakly modified. Although these effects are extremely small for realistic environments, radial pressure systematically amplifies the deviations from the vacuum geometry relative to the \ec model.

\subsection{Quasinormal modes}

We now investigate the effects of the anisotropic environment on the \qnm spectrum, which determines the ringdown signal of the perturbed \bh spacetime.
With the Fourier transform $\psi(t, r)=\int d\omega\,\psi(\omega, r) e^{-i \omega t}$, Eq.~\eqref{eq:AxialTime} can be rewritten in the frequency domain as
\begin{equation}
\begin{aligned}\label{eq:AxialFrequency}
\left(\frac{\partial^2}{\partial r_*^2}+\omega^2 -V^{\mathrm{sta/irr}}\right) \psi=0,
\end{aligned}
\end{equation}
where the eigenvalues $\omega$ are the corresponding \qnm{}s, defined by ingoing boundary conditions at the horizon and outgoing boundary conditions at spatial infinity:
\begin{equation}\label{eq:BC}
        \psi\rightarrow\begin{cases}
         \mathrm{e}^{-i \omega r_*}  & r\rightarrow 2M_{\mathrm{BH}}  \\
            \mathrm{e}^{+i \omega r_*} & r\rightarrow\infty
            \\\end{cases}.
\end{equation}

Following Ref.~\cite{Cardoso:2022whc}, we therefore expand Eq.~\eqref{eq:AxialFrequency} to linear order in the environmental compactness $\mathcal{C}$, yielding 
\begin{align}
    \label{eq:elambda}
    \frac{e^\lambda}{e^{\lambda_{\mathrm{vac}}}}&=1+\mathcal{O}(\mathcal{C}^2)\\
    \label{eq:enu}
    \frac{e^\nu}{e^{\nu_{\mathrm{vac}}}}&=1-2(1+\mathcal{K})\mathcal{C}+\mathcal{O}(\mathcal{C}^2)\\
    \frac{V^{\mathrm{sta/irr}}}{V^{\mathrm{RW}}}&=1-2(1+\mathcal{K})\mathcal{C}+\mathcal{O}(\mathcal{C}^2)
\end{align}
where the subscript ``vac'' denotes the vacuum ($\mathcal{C}=0$) case, and $V^{\mathrm{RW}}$ is the well-known \rw potential~\cite{Regge:1957td}, given by
\begin{equation}
    V^{\mathrm{RW}}=\left(1-\frac{2M_{\mathrm{BH}}}{r}\right)\left[\frac{\ell(\ell+1)}{r^2}-\frac{6 M_{\mathrm{BH}}}{r^3}\right].
\end{equation}
Based on this approximation, the corresponding \qnm{}s are expected to be redshifted as
\begin{equation}\label{eq:QNMs}
    \frac{\omega}{\omega_{\mathrm{vac}}}=1-\left(1+\mathcal{K}\right)\mathcal{C}+\mathcal{O}(\mathcal{C}^2).
\end{equation}
This result is also consistent with the geodesic analysis in Sec.~\ref{sec:Geodesic} through the well-known eikonal relation~\cite{Cardoso:2008bp}
\begin{equation}
\omega=\Omega_{\mathrm{LR}} \ell-i(n+1/2)|\lambda|,
\end{equation}
where $n$ is the overtone number.

\begin{figure*}
    \centering
\includegraphics[width=0.48\textwidth]{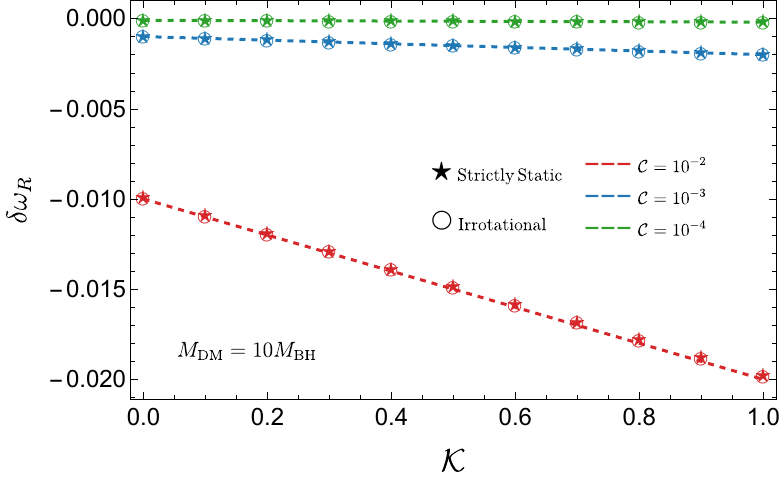}
\includegraphics[width=0.48\textwidth]{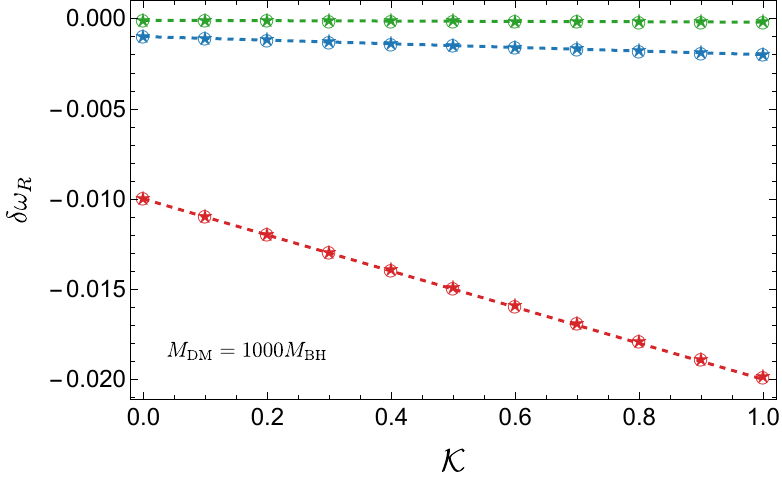}\\
\includegraphics[width=0.48\textwidth]{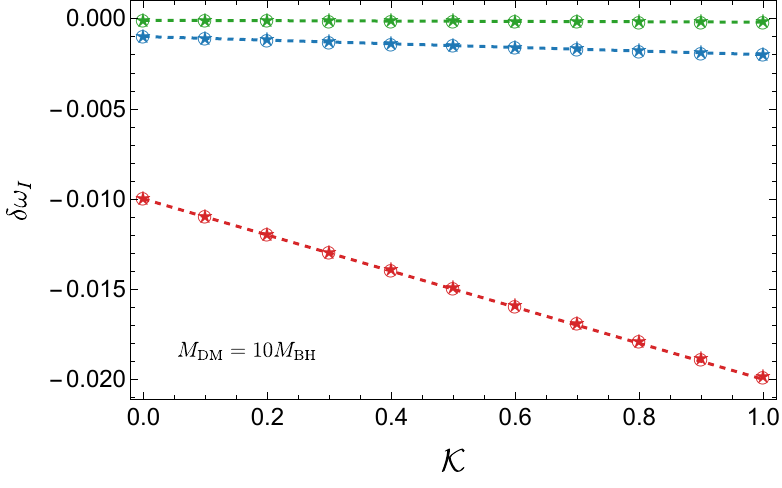}
\includegraphics[width=0.48\textwidth]{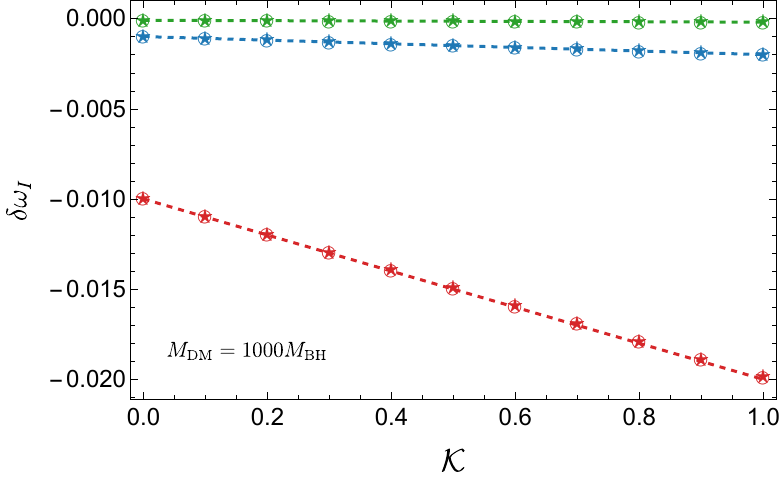}
   \caption{Relative deviations of the axial \qnm{}s with respect to the vacuum Schwarzschild values for the $(\ell,n)=(2,0)$ mode.
   The \qnm{}s are found to be sensitive to the compactness $\mathcal{C}=M_{\mathrm{DM}}/a_0$ and $\mathcal{K}$.
   The dashed lines correspond to the analytical prediction in Eq.~\eqref{eq:QNMs}, while the stars and circles denote the numerical results of the strictly static and irrotational cases, respectively.
   $\delta \omega_{R}$ denotes the relative deviation of the real parts, while $\delta \omega_{I}$ denotes the relative deviation of the imaginary parts.
   }
    \label{fig:QNMs}
\end{figure*}

By introducing the relative deviation
\begin{equation}\label{eq:deviation}
    \delta \omega=\frac{\omega-\omega_{\mathrm{vac}}}{\omega_{\mathrm{vac}}},
\end{equation}
we use the sixth-order WKB method to calculate the fundamental \qnm{}s with $(\ell,n)=(2,0)$~\cite{Iyer:1986np,Konoplya:2003ii,Matyjasek:2017psv}. 
As shown in Fig.~\ref{fig:QNMs}, the numerical results are sensitive to the compactness $\mathcal{C}$ and the sound speed $\mathcal{K}$, and agree very well with the analytical prediction in Eq.~\eqref{eq:QNMs}.

As discussed in Sec.~\ref{sec:Framework} and Appendix~\ref{app:Perturbation}, the perturbation equations, and hence the \qnm{} spectrum, depend in principle on the choice of fluid variables. Nevertheless, because the anisotropy parameter $\sigma$ is small in realistic astrophysical environments (cf. Fig.~\ref{fig:sigma} and~\ref{fig:energy}), both the strictly static and irrotational prescriptions produce the same leading-order correction. This indicates a certain universality of the axial \qnm{} response, which is governed predominantly by the background geometry rather than by the detailed dynamics of matter perturbations.

The leading environmental effect is therefore still an overall gravitational redshift. Compared to the \ec model~\cite{Cardoso:2021wlq}, radial pressure systematically amplifies this correction and can enhance the deviation from the vacuum spectrum by up to an $\mathcal{O}(1)$ factor within the parameter range allowed by Eqs.~\eqref{eq:K14} and~\eqref{eq:SoundSpeed}.

We stress that this redshift is not specific to the polytropic model. For a generic weak environment, the expansion of Appendix~\ref{app:Limits} gives $e^{\nu}=(1-2M_{\mathrm{BH}}/r)(1+\epsilon H)$ at first order, with $H$ determined by Eq.~\eqref{eq:weakTOV}. If the matter is distributed on scales much larger than $M_{\mathrm{BH}}$, $H$ is approximately constant near the light ring, $\epsilon H\simeq-2\int 4\pi r(\rho+p_r)\,dr$, which reduces to $-2(1+\mathcal{K})\mathcal{C}$ for our model. Therefore, the same redshift is expected for any \eos and density profile, as long as the environment is dilute and much more extended than the \bh.

\subsection{Tidal Love numbers}

Beyond the dynamical response encoded in the \qnm spectrum, it is also interesting to investigate the tidal response of the system, characterized by the tidal Love numbers.
Under the strictly static and irrotational constraints, respectively, together with $\dot{h}_i=0$, Eq.~\eqref{eq:axial3} yields
\begin{equation}
\begin{aligned}
    &e^{-\lambda} h_0^{\prime \prime}-4 \pi r(\rho+p_r) h_0^{\prime}\\
    &-\left[\frac{\ell(\ell+1)}{r^2}-\frac{4 M}{r^3}+8 \pi(\rho+p_r-2\sigma)\right] h_0=0 ,
\end{aligned}
\end{equation}
for the strictly static case, while for the irrotational case,
\begin{equation}
\begin{aligned}
& e^{-\lambda}h_0^{\prime \prime}-4 \pi r(\rho+p_r) h_0^{\prime}\\
& -\left[\frac{\ell(\ell+1)}{r^2}-\frac{4 M}{r^3}-8 \pi(\rho+p_r)\right] h_0=0.
\end{aligned}
\end{equation}

As discussed in~\cite{Pani:2018inf}, the definition and normalization of magnetic Love numbers in environmental spacetimes are subtle and prescription dependent.
To facilitate direct comparison with previous work, we adopt the definition and extraction procedure of~\cite{Cardoso:2017cfl,Cardoso:2019upw,Chakraborty:2024gcr} (see also~\cite{DOnofrio:2026ulh} for a discussion and alternative procedure). The corresponding magnetic Love numbers are given by
\begin{align}
    k_{\ell=2}^{\mathrm{B (sta)}}&=\frac{M_{\mathrm{DM}}a_0^4\left[5-17\mathcal{K}+12\left(1-3\mathcal{K}\right)\ln\left(a_0/\mathcal{L}\right)\right]}{3\mathcal{L}^5},\\
    k_{\ell=2}^{\mathrm{B (irr)}}&=\frac{M_{\mathrm{DM}}a_0^4\left[1+\mathcal{K}-4\left(1-\mathcal{K}\right)\ln\left(a_0/\mathcal{L}\right)\right]}{3\mathcal{L}^5},
\end{align}
For $\mathcal{K}=0$, these reduce to the ``up'' and ``down'' solutions of~\cite{Chakraborty:2024gcr}, respectively.

Here, $\mathcal{L}$ is the characteristic length scale associated with normalization.
Choosing $\mathcal{L}=M_{\mathrm{BH}}+M_{\mathrm{DM}}$, the strictly static Love number reduces to the result of~\cite{Cardoso:2021wlq}.
However, as discussed in~\cite{Chakraborty:2024gcr}, a more natural choice is $\mathcal{L}=a_0$.
In the realistic regime with $a_0\gg M_{\mathrm{DM}},\,M_{\mathrm{BH}}$, the corresponding Love numbers become
\begin{align}
    k_{\ell=2}^{\mathrm{B (sta)}}&=\frac{\left(5-17\mathcal{K}\right)}{3}\mathcal{C},\\
    k_{\ell=2}^{\mathrm{B (irr)}}&=\frac{\left(1+\mathcal{K}\right)}{3}\mathcal{C}.
\end{align}
The relative deviation between the two cases is given by
\begin{equation}
    \Delta  k_{\ell=2}^{\mathrm{B}}=\frac{ k_{\ell=2}^{\mathrm{B (irr)}}- k_{\ell=2}^{\mathrm{B (sta)}}}{ k_{\ell=2}^{\mathrm{B (irr)}}}=18-\frac{22}{1+\mathcal{K}}.
\end{equation}
Interestingly, the strictly static Love number vanishes at $\mathcal{K}=5/17$ and becomes negative for larger $\mathcal{K}$, whereas the irrotational Love number remains positive within the theoretically allowed range.
Moreover, increasing $\mathcal{K}$ enhances the difference between the two cases, while the Love numbers coincide at $\mathcal{K}=2/9$.
This behavior indicates that the tidal response is significantly more sensitive to the radial pressure than the geodesic structure and \qnm{}s.
In particular, within the theoretically allowed parameter range constrained by Eq.~\eqref{eq:SoundSpeed}, the inclusion of radial pressure can lead to order-unity modifications of the strictly static magnetic Love number relative to the \ec case.
Given their large magnitude, environmental tidal effects in \bh binaries are expected to be detectable by future observations~\cite{DeLuca:2021ite,DeLuca:2022xlz,Cannizzaro:2024fpz} (see~\cite{Chakraborty:2026qru} for a recent overview).
We stress that the above comparison concerns the relative sensitivity to the properties of the environment. Unlike the \qnm{} shift, which at leading order is degenerate with a redshift of the \bh mass~\cite{Spieksma:2024voy}, the tidal Love numbers vanish in vacuum, so any nonzero value would be a deviation from the standard vacuum BH case.

\section{Discussion}

In this work, we developed a relativistic framework describing \bh{}s embedded in anisotropic astrophysical environments with nonvanishing radial pressure, extending previous studies based on \ec and isotropic matter configurations~\cite{Cardoso:2021wlq,Cardoso:2022whc,Chakraborty:2024gcr,Kojima:1992ie,Zhao:2023tyo}. Within this framework, we derived a class of closed-form solutions supported by a polytropic radial pressure and generic pressure anisotropy, and investigated their signatures in terms of the geodesic structure, axial \qnm{}s, and tidal response of the \bh.

A first important result concerns the physical viability of the model. We found that the \dec{} is violated in the innermost region surrounding the \bh, thereby constraining the allowed parameter space. Interestingly, realistic \dm spike models predict a substantial depletion of matter within $r\sim4M_{\mathrm{BH}}$ or near the \isco~\cite{Gondolo:1999ef,Sadeghian:2013laa,Speeney:2022ryg,Mitra:2025tag,Zhao:2023tyo}, approximately coinciding with the region where the \dec{} is violated. This suggests that the effective-fluid description is likely to break down precisely where these violations occur and motivates the development of more realistic models incorporating explicit spike profiles.

Our analysis reveals a clear hierarchy in the sensitivity of different observables to the surrounding environment. Consistent with previous studies of \ec configurations~\cite{Cardoso:2021wlq}, both the geodesic structure and the axial \qnm spectrum are found to be largely governed by an overall gravitational redshift induced by the environment. The presence of radial pressure systematically enhances these effects, effectively replacing the leading-order redshift parameter $\mathcal{C}=M_{\mathrm{DM}}/a_0$ with $(1+\mathcal{K})\mathcal{C}$, where $\mathcal{K}$ can be interpreted as the squared effective radial sound speed. Nevertheless, for realistic astrophysical environments, the resulting corrections remain small.

In contrast, the tidal Love numbers are substantially more sensitive to anisotropy and can exhibit order-unity deviations throughout the physically allowed parameter range. In particular, radial pressure can induce corrections that are comparable to, or even larger than, those associated with the matter distribution itself. These results suggest that tidal observables provide a significantly more powerful probe of environmental effects~\cite{DeLuca:2021ite,DeLuca:2022xlz,Cannizzaro:2024fpz} than either geodesic or ringdown measurements, and highlight the importance of including pressure effects when modeling realistic environments surrounding \bh{}s.

An additional feature of anisotropic environments is that the axial perturbation equations are no longer automatically closed because of the extra fluid degrees of freedom associated with the anisotropic sector. In this work, we considered two representative closure prescriptions, namely the strictly static and irrotational cases. Despite their different treatment of matter perturbations, both prescriptions yield the same leading-order correction to the \qnm spectrum in realistic environments, suggesting a certain degree of universality in the dynamical response. The tidal sector exhibits a richer behavior: magnetic Love numbers can become strongly suppressed, vanish, or even change sign for sufficiently large anisotropies and radial sound speed. Furthermore, although the perturbation equations smoothly reduce to the isotropic limit as $\sigma\to0$, the distinction between strictly static and irrotational magnetic Love numbers persists, consistent with previous studies of relativistic tidal responses~\cite{Pani:2018inf}. These results indicate that the perturbative dynamics may depend sensitively on the microscopic realization of the effective fluid and motivate a deeper investigation of the physical interpretation and consistency of different closure prescriptions.

The present analysis focused on axial perturbations, for which the system can still be reduced to a single master equation once a closure condition is specified. A complete understanding of anisotropic environments requires extending the analysis to the polar sector, where matter fluctuations couple dynamically to the spacetime geometry and are expected to produce a significantly richer phenomenology~\cite{Cardoso:2022whc,Speeney:2024mas,Mitra:2025tag}.
For instance, time-domain studies of the \ec found that, besides a redshifted prompt ringdown, the coupling to matter excites long-lived fluid-driven modes~\cite{Cardoso:2022whc,Spieksma:2024voy}, while the polar \gw fluxes are not captured by a simple redshift~\cite{Speeney:2024mas}.
For completeness, we provide the corresponding perturbation equations in~\cite{MyGithub}, leaving their detailed analysis to future work.

Several other extensions are worth pursuing. Rotating \bh{}s constitute a particularly interesting direction, since the matter distribution becomes more concentrated near the horizon and may therefore lead to stronger environmental signatures~\cite{Fernandes:2025osu,Yue:2026gju}. Additional developments include time-domain evolutions and more realistic environmental models based on self-consistent \dm spike profiles~\cite{Speeney:2022ryg}. More broadly, the framework developed here provides a useful starting point for systematic studies of environmental effects in \gw astrophysics, including ringdown spectroscopy, \emri waveform modeling, and future precision tests of strong-field gravity.

\begin{acknowledgments}
This work is supported by the MUR FIS2 Advanced Grant ET-NOW (CUP:~B53C25001080001) and the INFN TEONGRAV initiative.
Y.Z. acknowledges the PhD scholarship from the China Scholarship Council (CSC). 
\end{acknowledgments}

\appendix

\section{Perturbation Equations}\label{app:Perturbation}

In this appendix, we present the perturbation decomposition and summarize the corresponding equations.
The complete expressions can be found in the \texttt{Mathematica} notebook~\cite{MyGithub}.

\subsection{Decomposition of perturbations}

With the adoption of the \rw gauge~\cite{Regge:1957td,Nagar:2005ea}, the perturbed metric of Eq.~\eqref{eq:gmunuExpand} can be split into axial (odd-parity) and polar (even-parity) sectors as
\begin{equation}
g_{\mu\nu}^{\mathrm{axial}(1)}=\sum_{\ell=2}^{\infty} \sum_{m=-\ell}^{\ell}\left(\begin{array}{cccc}
0 & 0 & h_0^{\ell m} S_{\theta}^{\ell m}  & h_0^{\ell m} S_{\phi}^{\ell m} \\
* & 0 & h_1^{\ell m} S_{\theta}^{\ell m}  & h_1^{\ell m} S_{\phi}^{\ell m} \\
* & * & 0 & 0 \\
* & * & * & 0
\end{array}\right),
\end{equation}
and
\begin{equation}\begin{aligned}
&g_{\mu\nu}^{\mathrm{polar}(1)}=\sum_{\ell=2}^{\infty} \sum_{m=-\ell}^{\ell}Y^{\ell m}\times\\
&\left(\begin{array}{cccc}
e^\nu H_0^{\ell m} & H_1^{\ell m} & 0 & 0 \\
* & e^\lambda H_2^{\ell m} & 0 & 0 \\
* & * & r^2 K^{\ell m} & 0 \\
* & * & * & r^2 \sin^2\theta K^{\ell m}
\end{array}\right),
\end{aligned}\end{equation}
where the asterisks represent symmetric components, $Y^{\ell m}=Y^{\ell m}(\theta,\phi)$ are the scalar spherical harmonics, and $\left(S^{\ell m} _\theta,S^{\ell m} _\phi\right)=\left(-\csc\theta Y_{, \phi}^{\ell m} , \sin \theta Y_{, \theta}^{\ell m} \right)$ are the (axial parity) vector spherical harmonics.
$\left(h_0^{\ell m},h_1^{\ell m}\right)$ and $\left(H_0^{\ell m},H_1^{\ell m},H_2^{\ell m},K^{\ell m}\right)$ are functions of $(t,r)$ and describe the metric perturbations in the axial and polar sectors, respectively.

We expand the perturbations of the density and pressures in scalar spherical harmonics as
\begin{equation}
p_r^{(1)}(t, r, \theta, \phi)=\sum_{\ell=2}^{\infty} \sum_{m=-\ell}^{\ell} \delta p^{\ell m}_r(t, r) Y^{\ell m}(\theta, \phi),
\end{equation}
\begin{equation}
p_t^{(1)}(t, r, \theta, \phi)=\sum_{\ell=2}^{\infty} \sum_{m=-\ell}^{\ell} \delta p_t^{\ell m}(t, r) Y^{\ell m}(\theta, \phi),
\end{equation}
\begin{equation}
\rho^{(1)}(t, r, \theta, \phi)=\sum_{\ell=2}^{\infty} \sum_{m=-\ell}^{\ell} \delta \rho^{\ell m}(t, r) Y^{\ell m}(\theta, \phi).
\end{equation}

Additionally, with the constraint of $u_\mu u^\mu=-1$, the perturbation of the four-velocity can be described by only three variables $\left(R^{\ell m}, V^{\ell m}, U^{\ell m}\right)$, and expanded in scalar and vector spherical harmonics as~\cite{Kojima:1992ie}
\begin{equation}
u^{t(1)}=\frac{1}{2\sqrt{e^{\nu}}} \sum_{\ell=2}^{\infty} \sum_{m=-\ell}^{\ell} H_0^{\ell m} Y^{\ell m},
\end{equation}
\begin{equation}
u^{r(1)}=\frac{1}{ e^{\lambda}\sqrt{e^{\nu}}(\rho+p_r)} \sum_{\ell=2}^{\infty} \sum_{m=-\ell}^{\ell} R^{\ell m} Y^{\ell m},
\end{equation}
\begin{equation}
\begin{aligned}
u^{\theta(1)}=&\frac{1}{ \sqrt{e^{\nu}}(\rho+p_t) r^2}\times\\
&\sum_{\ell=2}^{\infty} \sum_{m=-\ell}^{\ell}\left(V^{\ell m} Y^{\ell m}_{,\theta}-U^{\ell m}\csc\theta Y^{\ell m}_{,\phi}\right) ,
\end{aligned}
\end{equation}
\begin{equation}
\begin{aligned}
u^{\phi(1)}=&\frac{1}{\sqrt{e^{\nu}} (\rho+p_t) r^2 \sin ^2 \theta}\times \\
&\sum_{\ell=2}^{\infty} \sum_{m=-\ell}^{\ell}\left(V^{\ell m} Y^{\ell m}_{,\phi}+U^{\ell m}\sin\theta Y^{\ell m}_{,\theta}\right) .
\end{aligned}
\end{equation}
The perturbation of the unit radial vector is constrained by $k_\mu k^\mu=1$ and $u_\mu k^\mu=0$, which provide two constraint equations for $ k^{t(1)}$ and $k^{r(1)}$, and two additional variables $(Z^{\ell m},X^{\ell m})$ are required, leading to~\cite{DOnofrio:2026ulh}
\begin{equation}
\begin{aligned}
    k^{t(1)}=&\frac{1}{e^\nu\sqrt{e^{\lambda}}}\sum_{\ell=2}^{\infty} \sum_{m=-\ell}^{\ell}\left[ H_1^{\ell m}+ \frac{R^{\ell m}}{(\rho+p_r)}\right]Y^{\ell m},
\end{aligned}
\end{equation}
\begin{equation}
    k^{r(1)}=-\frac{1}{2\sqrt{e^{\lambda}}}\sum_{\ell=2}^{\infty} \sum_{m=-\ell}^{\ell}H_2^{\ell m} Y^{\ell m},
\end{equation}
\begin{equation}
\begin{aligned}
    k^{\theta(1)}=&\frac{1}{\sqrt{e^{\lambda}} r^2 }\times \\
    &\sum_{\ell=2}^{\infty} \sum_{m=-\ell}^{\ell} \left(Z^{\ell m} Y^{\ell m}_{,\theta}-X^{\ell m}\csc\theta Y^{\ell m}_{,\phi}\right) ,
\end{aligned}
\end{equation}
\begin{equation}
\begin{aligned}
    k^{\phi(1)}=&\frac{1}{\sqrt{e^{\lambda}}  r^2 \sin^2\theta}\times \\
    &\sum_{\ell=2}^{\infty} \sum_{m=-\ell}^{\ell} \left(Z^{\ell m}  Y^{\ell m}_{,\phi}+X^{\ell m} \sin\theta Y^{\ell m}_{,\theta}\right) ,
\end{aligned}
\end{equation}
The fluid decomposition reduces to that of~\cite{Kojima:1992ie,DOnofrio:2026ulh} under the transformation
\begin{equation}\label{eq:transform}
    \left(R^{\ell m},U^{\ell m},V^{\ell m}\right)\rightarrow\left(\frac{e^\nu}{4\pi}R^{\ell m},\frac{e^\nu}{4\pi}U^{\ell m},\frac{e^\nu}{4\pi}V^{\ell m}\right),
\end{equation}
\begin{equation}
    X^{\ell m}\rightarrow\frac{1}{4\pi (\rho+p_t)}U_k^{\ell m},
\end{equation}
where $U_k^{\ell m}$ is the notation of~\cite{DOnofrio:2026ulh}.
The corresponding components of $T_{\mu\nu}^{\mathrm{env(1)}}$ are then given by
\begin{equation}
    T_{tt}^{\mathrm{env(1)}}=e^{\nu}\sum_{\ell=2}^{\infty} \sum_{m=-\ell}^{\ell} \left(\delta\rho^{\ell m}-\rho H_0^{\ell m}\right)Y^{\ell m},
\end{equation}
\begin{equation}
    T_{tr}^{\mathrm{env(1)}}=-\sum_{\ell=2}^{\infty} \sum_{m=-\ell}^{\ell}\left[  \rho H_1^{\ell m}+R^{\ell m}\right]Y^{\ell m},
\end{equation}
\begin{equation}
\begin{aligned}
    T_{t\theta}^{\mathrm{env(1)}}&=\sum_{\ell=2}^{\infty} \sum_{m=-\ell}^{\ell} \\
    &\left[\left( \rho  h_0^{\ell m}+U^{\ell m}\right)\left(\csc\theta Y^{\ell m}_{,\phi}\right)-V^{\ell m}Y^{\ell m}_{,\theta}\right],
\end{aligned}
\end{equation}
\begin{equation}
\begin{aligned}
    T_{t\phi}^{\mathrm{env(1)}}&=-\sum_{\ell=2}^{\infty} \sum_{m=-\ell}^{\ell}\\
    &\left[\left(\rho  h_0^{\ell m}+U^{\ell m}\right)\left(\sin \theta Y^{\ell m}_{,\theta}\right)+V^{\ell m} Y^{\ell m}_{,\phi}\right],
\end{aligned}
\end{equation}
\begin{equation}
    T_{rr}^{\mathrm{env(1)}}=e^{\lambda }\sum_{\ell=2}^{\infty} \sum_{m=-\ell}^{\ell}  (\delta p_r^{\ell m}+p_r H_2^{\ell m})Y^{\ell m},
\end{equation}
\begin{equation}
\begin{aligned}
    T_{r\theta}^{\mathrm{env(1)}}&=\sum_{\ell=2}^{\infty} \sum_{m=-\ell}^{\ell} \\
    &\left[\sigma Z^{\ell m}Y^{\ell m}_{,\theta}-\left(p_r h_1^{\ell m}+\sigma X^{\ell m}\right) \left(\csc \theta Y^{\ell m}_{,\phi}\right)\right],
\end{aligned}
\end{equation}
\begin{equation}
\begin{aligned}
    T_{r\phi}^{\mathrm{env(1)}}&=\sum_{\ell=2}^{\infty} \sum_{m=-\ell}^{\ell} \\
    &\left[\sigma Z^{\ell m}Y^{\ell m}_{,\phi}+\left(p_r h_1^{\ell m}+\sigma X^{\ell m}\right) \left(\sin \theta Y^{\ell m}_{,\theta}\right)\right],
\end{aligned}
\end{equation}
\begin{equation}
    T_{\theta\theta}^{\mathrm{env(1)}}=r^2\sum_{\ell=2}^{\infty} \sum_{m=-\ell}^{\ell} \left[\delta p_t^{\ell m}+p_t K^{\ell m}\right]Y^{\ell m},
\end{equation}
\begin{equation}
    T_{\theta\phi}^{\mathrm{env(1)}}=0,
\end{equation}
\begin{equation}
    T_{\phi\phi}^{\mathrm{env(1)}}=\sin ^2\theta T_{\theta\theta}^{\mathrm{env(1)}}.
\end{equation}
All these expressions reduce to Appendix~A of~\cite{Kojima:1992ie} with the variable transformation Eq.~\eqref{eq:transform} in the isotropic limit, namely, $p=p_r=p_t$, $\sigma=0$, and $\delta p^{\ell m}=\delta p_r^{\ell m}=\delta p_t^{\ell m}$.

\subsection{Axial sector}

In this section, we derive the axial perturbation equations from Eqs.~\eqref{eq:Einstein} and~\eqref{eq:DTmunu}; in the following expressions we omit the subscript ``$\ell m$''.
The components $(\mathcal{E}_{\theta\theta},\mathcal{E}_{\theta\phi},\mathcal{E}_{\phi\phi})$, $(\mathcal{E}_{r\theta},\mathcal{E}_{r\phi})$, and $(\mathcal{E}_{t\theta},\mathcal{E}_{t\phi})$ of the axial sector lead to the following equations, respectively:
\begin{equation}\label{eq:axial1}
e^{-\nu} \dot{h}_0-e^{-\lambda} h_1^{\prime}-\frac{1}{r^2}\left[2 M-4 \pi(\rho-p_r) r^3\right] h_1=0,
\end{equation}
\begin{equation}\label{eq:axial2}
\begin{aligned}
    &e^{-\nu}\left(\dot{h}_0^{\prime}-\ddot{h}_1\right)-\frac{2 e^{-\nu}}{r} \dot{h}_0-\left[\frac{(\ell-1)(\ell+2)}{r^2}-16\pi\sigma\right] h_1\\&
    +16\pi\sigma X=0,
\end{aligned}
\end{equation}
\begin{equation}\label{eq:axial3}
\begin{aligned}
& e^{-\lambda}\left(h_0^{\prime \prime}-\dot{h}_1^{\prime}\right)-4 \pi(\rho+p_r) r\left(h_0^{\prime}-\dot{h}_1\right)-\frac{2 e^{-\lambda}}{r} \dot{h}_1 \\
& -\frac{1}{r^3}\left[\ell(\ell+1) r-4 M+8 \pi(\rho+p_r-2\sigma) r^3\right] h_0\\
&-16\pi U=0,
\end{aligned}
\end{equation}
and the components $(\nabla_\mu T^{\mu \theta},\nabla_\mu T^{\mu \phi})$ result in
\begin{equation}\label{eq:axial4}
   \begin{aligned}
       &\partial_t\left[U+(\rho+p_t) h_0\right]+e^{\nu-\lambda}\sigma\partial_r\left(X+h_1\right)\\
       &+e^\nu\left\{\left[\frac{1+e^{-\lambda}}{r}-4\pi r\left(\rho-p_r\right)\right]\sigma+e^{-\lambda}\sigma'\right\}\left(X+h_1\right)\\
       &=0,
   \end{aligned}
\end{equation}
where the background \tov{} equations, Eqs.~\eqref{eq:tov1}--\eqref{eq:tov3}, were used to simplify the equations when necessary, and all the remaining components are $0$.
Eq.~\eqref{eq:axial1} implies~\cite{Pani:2018inf}
\begin{align}\label{eq:h0}
\dot{h}_0&=e^{(\nu-\lambda) / 2}(\psi r)^{\prime},\\
\label{eq:h1}
h_1&=e^{(\lambda-\nu) / 2} \psi r.
\end{align}
Substituting Eqs.~\eqref{eq:h0} and~\eqref{eq:h1} into Eq.~\eqref{eq:axial2}, we obtain the perturbation equations
\begin{equation}
\left(\frac{\partial^2}{\partial r_*^2} -\frac{\partial^2}{\partial t^2} -V^{\mathrm{axial}}\right) \psi=S^{\mathrm{axial}},
\end{equation}
\begin{equation}
    V^{\mathrm{axial}}=e^\nu\left[\frac{\ell(\ell+1)}{r^2}-\frac{6 M}{r^3}+4 \pi(\rho-p_r-4\sigma)\right],
\end{equation}
\begin{equation}\label{eq:source}
    S^{\mathrm{axial}}=-\frac{16 \pi  \sigma e^{(3 \nu-\lambda)/2} }{r}X
\end{equation}
where $r_*$ is the tortoise coordinate defined by
\begin{equation}
    \frac{\partial r_*}{\partial r}=e^{(\lambda-\nu)/2}.
\end{equation}

Eq.~\eqref{eq:axial4} provides a constraint to $X$.
To close this system and determine Eq.~\eqref{eq:source}, an additional condition is required.
Here, we focus on two physical cases, namely the \textit{strictly static} and \textit{irrotational} cases~\cite{Pani:2018inf,DOnofrio:2026ulh}, corresponding to the ``up'' and ``down'' prescriptions of~\cite{Chakraborty:2024gcr}.

The strictly static case requires~\cite{Chakraborty:2024gcr}
\begin{equation}
    U=X=0,
\end{equation}
which leads to
\begin{equation}\label{eq:static}
\left(\frac{\partial^2}{\partial r_*^2} -\frac{\partial^2}{\partial t^2} -V^{\mathrm{sta}}\right) \psi=0,
\end{equation}
\begin{equation}\label{eq:Vstatic}
    V^{\mathrm{sta}}=e^\nu\left[\frac{\ell(\ell+1)}{r^2}-\frac{6 M}{r^3}+4 \pi(\rho-p_r-4\sigma)\right].
\end{equation}

Additionally, the irrotational case requires vanishing vorticity of $u^\mu$~\cite{DOnofrio:2026ulh}
\begin{equation}
    \omega^\alpha\left[u^\gamma\right]=\frac{1}{2} \epsilon^{\alpha \beta \mu \nu} (\nabla_\mu u_{\beta}) u_\nu=0,
\end{equation}
which gives
\begin{equation}
    U=-(\rho+p_t)h_0.
\end{equation}
Substituting this condition, Eq.~\eqref{eq:axial4} yields a constraint for $X+h_1$, and the straightforward solution is
\begin{equation}
    X=-h_1.
\end{equation}
Coincidentally, this solution also satisfies the vanishing vorticity of $k^\mu$, which reads
\begin{equation}
    \omega^\alpha\left[k^\gamma\right]=\frac{1}{2} \epsilon^{\alpha \beta \mu \nu} (\nabla_\mu k_{\beta}) k_\nu=0,
\end{equation}
implying irrotationality for both $u^\mu$ and $k^\mu$, and a related discussion can be found in~\cite{DOnofrio:2026ulh}.
Therefore, the corresponding perturbation equation becomes
\begin{equation}\label{eq:irr}
\left(\frac{\partial^2}{\partial r_*^2} -\frac{\partial^2}{\partial t^2} -V^{\mathrm{irr}}\right) \psi=0,
\end{equation}
\begin{equation}\label{eq:Virr}
    V^{\mathrm{irr}}=e^\nu\left[\frac{\ell(\ell+1)}{r^2}-\frac{6 M}{r^3}+4 \pi(\rho-p_r)\right].
\end{equation}

\subsection{Polar sector}

The polar-sector equations are complicated by the coupling between metric and fluid perturbations, and we summarize them in the \texttt{Mathematica} notebook~\cite{MyGithub}.
The polar-sector component $\mathcal{E}_{\theta\phi}$ implies
\begin{equation}
    H_2^{\ell m}=H_0^{\ell m},
\end{equation}
which was used to simplify the equations.
Moreover, the \tov{} equations, Eqs.~\eqref{eq:tov1}--\eqref{eq:tov3}, were also used to simplify the equations.

\section{Particular cases}\label{app:Limits}

In this appendix, we show that our results include the vacuum, isotropic, \ec, and weak-environment limits as particular cases.
For definiteness, we focus on the axial sector.

We first consider the \textit{isotropic limit} with $p=p_r=p_t$, hence $\sigma=0$.
Our variables are related to~\cite{Kojima:1992ie,Pani:2018inf} through Eq.~\eqref{eq:transform}. The background equations, Eqs.~\eqref{eq:tov1}--\eqref{eq:tov3}, and perturbation equations, Eqs.~\eqref{eq:axial1}--\eqref{eq:axial3}, reduce to Eq.~(1) and Eqs.~(4)--(6) of~\cite{Pani:2018inf}, respectively.
The constraint Eq.~\eqref{eq:axial4} is reduced to
\begin{equation}
   \begin{aligned}
       &\partial_t\left[U+(\rho+p_t) h_0\right]=0,
   \end{aligned}
\end{equation}
which agrees with Eq.~(12) of~\cite{Pani:2018inf}.
The effective potentials of the strictly static and irrotational cases, given by Eqs.~\eqref{eq:Vstatic} and~\eqref{eq:Virr}, respectively, both become
\begin{equation}
    V^{\mathrm{iso}}=e^\nu\left[\frac{\ell(\ell+1)}{r^2}-\frac{6 M}{r^3}+4 \pi(\rho-p)\right],
\end{equation}
which coincides with Eq.~(11) of~\cite{Pani:2018inf}.

Furthermore, in the \textit{vacuum limit} with $\rho=p=0$, the \tov{} equations, Eqs.~\eqref{eq:tov1}--\eqref{eq:tov3}, imply that the background reduces to Schwarzschild spacetime, with $e^\nu=1-2M_{\mathrm{BH}}/r$ and $M=M_{\mathrm{BH}}$.
Thus, both cases lead to the well-known \rw equations with the effective potential given by~\cite{Regge:1957td}
\begin{equation}\label{eq:RW}
    V^{\mathrm{RW}}=\left(1-\frac{2M_{\mathrm{BH}}}{r}\right)\left[\frac{\ell(\ell+1)}{r^2}-\frac{6 M_{\mathrm{BH}}}{r^3}\right].
\end{equation}

In the \textit{\ec limit} with $p_r=0$ and $\sigma=-p_t$, Eq.~\eqref{eq:tov1} and Eq.~\eqref{eq:tov3} reduce to Eq.~(4) and Eq.~(10) of~\cite{Cardoso:2021wlq} respectively.
Additionally, Eq.~\eqref{eq:tov3} degenerates into a constraint given by
\begin{equation}
    p_t=\frac{M\rho}{2\left(r-2M\right)},
\end{equation}
which is Eq.~(5) of~\cite{Cardoso:2021wlq}.
Using the constraint on $p_t$ together with Eq.~\eqref{eq:tov1}, the effective potentials Eq.~\eqref{eq:Vstatic} and Eq.~\eqref{eq:Virr} can be rewritten as
\begin{equation}
\begin{aligned}
V_{\mathrm{EC}}^{\mathrm{sta}}=e^\nu\left[\frac{\ell(\ell+1)}{r^2}-\frac{6 M}{r^3}+ e^\lambda \frac{M^{\prime}}{r^2}\ \right],
\end{aligned}
\end{equation}
\begin{equation}
\begin{aligned}
V_{\mathrm{EC}}^{\mathrm{irr}}=e^\nu\left[\frac{\ell(\ell+1)}{r^2}-\frac{6 M}{r^3}+ \frac{M^{\prime}}{r^2}\ \right],
\end{aligned}
\end{equation}
where the irrotational case coincides with Eq.~(18) of~\cite{Cardoso:2021wlq}, and the strictly static case leads to an additional factor $e^{\lambda}=\left(1-2 M/r\right)^{-1}$.
Because the contribution of $e^\lambda$ is of order $\mathcal{O}(\mathcal{C}^2)$ in realistic environments, cf. Eq.~\eqref{eq:elambda}, these choices do not affect the leading-order analysis of~\cite{Cardoso:2021wlq}.

We now show that our results recover those of~\cite{Datta:2025ruh} with the \textit{weak-environment expansion}; the corresponding \texttt{Mathematica} notebook can be found at~\cite{MyGithub}.
Since the formalism in~\cite{Datta:2025ruh} is based on the strictly static case, Eqs.~\eqref{eq:static} and~\eqref{eq:Vstatic} should be considered at the perturbative level in comparison of the sourceless terms.
The environmental quantities are treated as small and tracked by $\epsilon$, namely
\begin{equation}
    \begin{aligned}
        &\rho=\epsilon\rho,\\
        &p_r=\epsilon p_r,\\
        &p_t=\epsilon p_t,
    \end{aligned}
\end{equation}
and the corresponding background metric functions are expanded as
\begin{equation}
    \begin{aligned}
        &e^{\nu(r)}=\left(1-\frac{2M_\mathrm{BH}}{r}\right)\left[1+\epsilon H(r)\right],\\
        &M(r)=\left[M_\mathrm{BH}+\epsilon m(r)\right],\\
        &e^{-\lambda(r)}=1-\frac{2 M(r)}{r}.
    \end{aligned}
\end{equation}
The \tov{} equations, Eqs.~\eqref{eq:tov1}--\eqref{eq:tov3}, give
\begin{equation}
    \begin{aligned}\label{eq:weakTOV}
        &m^{\prime}=4 \pi  r^2 \rho,\\
        &H^{\prime}=\frac{2 \left[m+4 \pi  r^2 (r-2 M_\mathrm{BH}) p_r\right]}{(r-2 M_\mathrm{BH})^2},\\
        &p_r^{\prime}=-\frac{M_\mathrm{BH} (p_r+\rho )}{r (r-2 M_\mathrm{BH})}-\frac{2 \sigma}{r},
    \end{aligned}
\end{equation}
which coincide with Eqs.~(18)--(19) of~\cite{Datta:2025ruh}.
To expand the perturbation equations and match the form of~\cite{Datta:2025ruh}, the master function should be redefined by
\begin{equation}
    \begin{aligned}
        &\phi(t,r)=-\sqrt{Z(r)}\psi(t,r),\\
        &Z(r)=1+\epsilon\left[\frac{H(r)}{2}-\frac{m(r)}{r f(r)}\right],
    \end{aligned}
\end{equation}
with $f(r)=1-2 M_\mathrm{BH}/r$.
Moreover, $\phi$ should be expanded as
\begin{equation}
    \phi=\phi_0+\epsilon\phi_1,
\end{equation}
where $\left(\phi_0,\phi_1\right)$ correspond to $\left(\phi_{\ell m}^{(1,0)},\phi_{\ell m}^{(1,1)}\right)$ of~\cite{Datta:2025ruh}.
The perturbation equations, Eqs.~\eqref{eq:static} and~\eqref{eq:Vstatic} then become
\begin{equation}
    \left(\frac{\partial^2}{\partial r_*^2} -\frac{\partial^2}{\partial t^2} -V^{\mathrm{RW}}\right) \phi_0=0,
\end{equation}
\begin{equation}
    \left(\frac{\partial^2}{\partial r_*^2} -\frac{\partial^2}{\partial t^2} -V^{\mathrm{RW}}\right) \phi_1=S_{1},
\end{equation}
where $V^{\mathrm{RW}}$ is the \rw potential given by Eq.~\eqref{eq:RW}, and the tortoise coordinate is also redefined as
\begin{equation}
    \frac{\partial r_*}{\partial r}=\frac{1}{f}.
\end{equation}
The source term $S_1$ is given by
\begin{equation}
    \begin{aligned}
        &S_1=\left[\frac{2 m}{r-2M_{\mathrm{BH}}}-H\right] \ddot{\phi}_0\\
        &-\left\{-\frac{2 [\ell (\ell+1) (r-2M_{\mathrm{BH}})-4 r+7M_{\mathrm{BH}}] }{(r-2M_{\mathrm{BH}}) r^3}m\right.\\
        &+\pi  \left(22-\frac{42M_{\mathrm{BH}}}{r}\right) p_r-\frac{20 \pi  (r-2M_{\mathrm{BH}}) }{r}p_t\\
        &\left.+2 \pi  (r-2M_{\mathrm{BH}}) \rho^{\prime}+6 \pi  \left(\frac{M_{\mathrm{BH}}}{r}-1\right) \rho\right\}\phi_0,
    \end{aligned}
\end{equation}
which coincide with the sourceless version of Eqs.~(47)--(49) in~\cite{Datta:2025ruh}.
Therefore, our results recover the vacuum, isotropic, \ec, and weak-environment limits under the appropriate approximations.

\bibliography{reference}

\end{document}